\documentclass[10pt,twocolumn]{article}
\usepackage[T1]{fontenc}
\usepackage[utf8]{inputenc}
\usepackage[margin=0.75in,columnsep=0.25in]{geometry}
\usepackage{amsmath,amssymb}
\usepackage{booktabs}
\usepackage{graphicx}
\usepackage{tabularx}
\usepackage{array}
\usepackage{pdflscape}
\usepackage[numbers,super,comma,sort&compress]{natbib}
\usepackage{tikz}
\usetikzlibrary{positioning,arrows.meta,calc}
\usepackage{microtype}
\usepackage[hidelinks]{hyperref}
\usepackage{url}

\newcommand{\hd}[1]{\begin{tabular}[b]{@{}c@{}}#1\end{tabular}}
\newcolumntype{L}[1]{>{\raggedright\arraybackslash}p{#1}}
\newcolumntype{Y}{>{\raggedright\arraybackslash}X}
\newcommand{\cum}{^{\mathrm{cum}}}

\begin{document}

\twocolumn[{%
\begin{center}
{\LARGE\bfseries Risk-set reconstruction balance: zero-targeted diagnostics
for artificial censoring in clone--censor--weight analyses\par}
\vspace{1em}
{\normalsize Yuta Nonomiya$^{1,2}$\\[0.5em]
$^{1}$Graduate School of Informatics, Osaka Metropolitan University\\
$^{2}$German Research Center for Artificial Intelligence GmbH (DFKI)\par}
\vspace{0.6em}
{\small \textbf{Correspondence:} Yuta Nonomiya, e-mail:
\href{mailto:nonomiya.yuta@omu.ac.jp}{nonomiya.yuta@omu.ac.jp}\par}
\end{center}
\vspace{0.4em}
\begin{center}
\begin{minipage}{0.9\textwidth}
\small
\noindent\textbf{Abstract.} In clone--censor--weight (CCW) analyses, inverse probability of censoring
weighting corrects selection from artificial censoring at protocol
deviation. Balance diagnostics, such as the end-of-grace
between-strategy standardised mean difference (SMD) and censoring balance,
compare different populations, and their zero targets under a correct
model are not formally stated. We show that, under positivity and a
correct censoring model, the weighted retained individuals reconstruct the
risk set before each censoring interval and the baseline population, and
we propose risk-set reconstruction balance (RRB), which measures how far
they depart from these two populations. Interval RRB has a zero target for
every history function, including post-baseline covariates, for which
between-strategy SMDs have none. Cumulative RRB keeps a zero target for
baseline covariates after outcome events, whereas the end-of-grace SMD
acquires a depletion term. In simulations, interval RRB responded to an
omitted term's effect on censoring whether or not the term was prognostic.
Under correct models, the exceedance rate of its studentised maximum over
a null reference for chance imbalance was 3.5--6.0\% (nominal 5\%) when
the censoring model pooled nine intervals and mostly 1.0--4.5\% under
censoring in a single interval. With many early events, cumulative RRB
stayed near zero, whereas the end-of-grace SMD reached 0.40; under loss to
follow-up that the weights ignored, cumulative RRB departed from zero and
interval RRB stayed small. In a case study, interval RRB flagged a
dispensing indicator outside the censoring model. Interval RRB checks the
censoring model, and cumulative RRB the representation of the baseline
population.

\end{minipage}
\end{center}
\vspace{1.5em}
}]

\section{Introduction}
Target trial emulation specifies the protocol of a hypothetical randomised
trial and emulates it with observational data
\cite{hernan2016a,hernan2016b}. If the initiator group of a treatment is
defined at initiation and the comparison group at eligibility, the
initiators are conditioned on survival to initiation, which induces
immortal-time bias \cite{suissa2008,hernan2016b}. The question can then be
recast as initiating within a grace period, for example 30 days from
baseline, against not initiating. In the clone--censor--weight (CCW) method
\cite{hernan2018}, every individual is followed under all strategies
compatible with the individual's baseline data, and artificial censoring
ends each of these follow-ups at the first deviation from the allocated
strategy. Inverse probability of censoring weighting (IPCW) corrects the
resulting selection \cite{robins2000a,cain2010}. As the weights are
estimated from a model for artificial censoring, whether they actually
achieve this correction should be assessed \cite{gaber2024}, and covariate
balance is widely used for this assessment.

A common diagnostic is the between-strategy standardised mean difference
(SMD) of baseline covariates at the end of the grace period
\cite{gaber2024,maringe2020,buranupakorn2021,cars2026}, which a CCW primer
describes as a simplified approach that excludes individuals with an event
before that time \cite{gaber2024}. Jackson proposes covariate balance
between censored and retained individuals (censoring balance), which is
defined for loss to follow-up and for artificial censoring at protocol
deviation and is applied in a CCW analysis of a randomised
trial \cite{jackson2019}.
The balance tables of the cobalt and WeightIt R packages compare the
weighted retained individuals with the unweighted sample at risk
\cite{greifer2026cobalt,greifer2026weightit}, and the ccwr R package
summarises the weights without covariate balance \cite{park2026ccwr}. A
systematic review of 113 CCW studies reports that 42 assessed covariate
balance after weighting \cite{manzanilla2026}, and one applied CCW study
that assesses balance states that the follow-up time at which balance
should be evaluated is an unanswered question \cite{cars2026}.

These diagnostics compare different populations, namely censored with
retained individuals (censoring balance), one strategy with another
(end-of-grace SMD), and the weighted retained individuals with the unweighted sample at
risk (cobalt and WeightIt). They draw on a broader literature on balance
between exposure groups after propensity-score matching or weighting, in
which imbalance reflects confounding of the exposure. This literature
includes balance diagnostics for time-varying exposures
\cite{jackson2016,adenyo2024}, covariate balancing weights for time-varying
treatments \cite{imai2015} and SMD conventions \cite{austin2009,austin2015,stuart2010}.
In CCW, the imbalance of interest arises from selection by artificial
censoring within each strategy, so a diagnostic must also specify the
population that the weighted retained individuals should reproduce.

The comparisons also differ in their zero targets. Jackson notes that
weights that produce independent censoring with respect to measured
covariates should remove the imbalance between censored and retained
individuals \cite{jackson2019}. Calibrated weights for loss to follow-up
\cite{yiu2022} and stable balancing weights for nonresponse
\cite{zubizarreta2015} make the weighted retained individuals reproduce a
reference sample, as the interval target below does. For the end-of-grace
SMD \cite{gaber2024}, the conditions under which a correct censoring model
implies a difference of zero have not been stated in the CCW setting. The
balance tables of cobalt and WeightIt
\cite{greifer2026cobalt,greifer2026weightit} take the unweighted sample at
risk as the target, which coincides with the interval target in general only in
the first interval with censoring. Between strategies, the individuals at risk are the same at baseline, so the
imbalance of baseline covariates after weighting arises from censoring and
weighting alone, at least until outcome events deplete the strategies
differently. Post-baseline covariates are affected by the treatment each
strategy prescribes, so their between-strategy differences can be
treatment effects. What has not
been organised in CCW is the relation between each diagnostic, its
reference population and its zero target, including how the weighted
retained individuals, together with those who had the outcome, reconstruct
the baseline population and how the diagnostics relate to between-strategy
SMDs. A reference for the imbalance that chance alone produces is also
lacking.

This paper makes three contributions. First, we show that, under positivity
and a correct censoring model, the weighted individuals retained under a
strategy reproduce the previously weighted risk set just before each
censoring interval (interval target) and, together with the individuals who
had the outcome, the baseline population (cumulative target). Censoring
balance then has a zero target, and the end-of-grace SMD acquires a
depletion term once outcome events have occurred. Second, we propose
risk-set reconstruction balance (RRB), which compares the weighted retained individuals with each target
population; interval RRB has a zero target for every function of the
covariate history, and cumulative RRB keeps a zero target for baseline
covariates after outcome events. Third, we provide a null reference for
chance imbalance in the maximum of interval RRB over strata.

Section~\ref{sec:notation} introduces the setting and notation.
Section~\ref{sec:targets} derives the interval and cumulative targets and
defines RRB, and Section~\ref{sec:null} constructs the null reference.
Sections~\ref{sec:sim} and~\ref{sec:simres} evaluate the diagnostics in a
three-part ADEMP simulation study \cite{morris2019}, and
Section~\ref{sec:emp} applies them to a public synthetic Medicare cohort.
Section~\ref{sec:disc} discusses the findings, their relation to previous
work and the limitations, and gives recommendations.

\section{Setting and notation}
\label{sec:notation}
We follow $n$ individuals over the intervals
$\mathcal I_k=[t_k,t_{k+1})$, $k=0,\dots,K-1$; an outcome or artificial
censoring during an interval carries its index. Individual $i$ has
baseline covariates $X_i$ and, for $\mathcal I_k$, time-varying covariates
$L_{ik}$, treatment $A_{ik}$ and the binary outcome indicator $Y_{ik}$. If individual
$i$ has the outcome, it occurs in $\mathcal I_{\tau_i}$. The history
$H_{ik}$ of individual $i$ comprises the covariates up to $\mathcal I_k$ and the
treatment before it.
Within an interval, covariates, treatment, artificial censoring, loss to
follow-up and the outcome occur in this order. A strategy $g$ is a
treatment rule, for example ``initiate within a window that ends with a
grace period'' or ``do not initiate before a given interval''
\cite{cain2010,wanis2024}. We consider strategies of treatment initiation,
in which initiation is absorbing; once an individual initiates,
$A_{ik'}=1$ for every later interval $k'$.

The risk-set indicator $R^g_{ik}$ equals 1 if, at the start of
$\mathcal I_k$, the treatment history of individual $i$ has remained
compatible with $g$, the outcome has not occurred and follow-up has not
been lost. The artificial-censoring indicator $D^g_{ik}$ equals 1 if
$R^g_{ik}=1$ and the treatment observed in $\mathcal I_k$ is incompatible
with $g$ \cite{hernan2018,cain2010,gaber2024b}; such an individual is absent from the risk set of the next
interval. For strategy $g$ and interval $\mathcal I_k$, the individuals at
risk form $\mathcal R=\{i:R^g_{ik}=1\}$, the censored individuals
$\mathcal C=\{i\in\mathcal R:D^g_{ik}=1\}$ and the retained individuals
$\mathcal U=\mathcal R\setminus\mathcal C$. We drop $i$ and $g$ where no ambiguity arises, writing,
for example, $D_k$ for $D^g_{ik}$, and keep $g$ and $g'$ only where two
strategies are compared. Censoring occurs in two phases
\cite{maringe2020,gaber2024}. An individual who initiates before the
window opens (under no initiation, before the given interval) is censored
at initiation, and an individual who has not initiated by the last interval
of the window is censored under the initiation strategy. Artificial
censoring of an individual at risk is a deterministic function of $A_k$, so
$H_k$ excludes $A_k$, as conditioning on it would make the probability of
censoring degenerate. In the CCW literature the record
of individual $i$ under strategy $g$ is called a clone; we keep the word
only in the name of the method.

Let $\pi_k=P(D_k=0\mid R_k=1,H_k)$ be the probability of
remaining uncensored given the history, with $\pi_k=1$ where censoring
cannot occur. Let $q_k=P(D_k=0\mid R_k=1)$,
which is common to all individuals at risk in the strategy-by-interval
stratum, hereafter stratum. The estimand is the risk
$P(Y^g_k=1\text{ for some }k<K)$ under each strategy, where $Y^g_k$ is the
outcome indicator under the strategy. It is estimated by a weighted Kaplan--Meier
estimator \cite{robins2000a,maringe2020} with the stabilised cumulative
weight
\[
W_{ik}=\prod_{l\le k}\frac{q_l}{\pi_{il}}=Q_kW^u_{ik},
\qquad Q_k=\prod_{l\le k}q_l,
\]
where $W^u_{ik}=\prod_{l\le k}1/\pi_{il}$ is the unstabilised weight
\cite{robins2000a,hernan2000,cole2008}, and $W_{i,-1}=1$. For the marginal risks
estimated here, the numerator must not depend on the history; a numerator
that depends on baseline covariates requires an analysis conditional on
them \cite{hernan2000,cole2008}. A common $q_k$
cancels within a stratum, so without truncation stabilisation affects only
the scale of the weights. In the analysis, $\pi$ is replaced by the fitted
value $\hat\pi$ of a pooled logistic model \cite{hernan2000,gaber2024b}, and
$q$ by the observed proportion remaining uncensored in the stratum, which
keeps the same symbol.
Weight concentration is summarised by stratum with the effective sample
size of the retained individuals,
$\mathrm{ESS}=(\sum_{\mathcal U}W_{ik})^2/\sum_{\mathcal U}W_{ik}^2$
\cite{kish1965}, divided by the number at risk $|\mathcal R|$, and with
percentiles of their weights, such as the 99th (W99)
\cite{cole2008,petersen2012}.

Diagnostics are computed for history functions $f_{ijk}=f_j(H_{ik})$,
such as covariates, transformations ($L_k^2$, $L_kL_{k-1}$,
$L_k-L_{k-1}$, cumulative means and indicators) and interactions. A function of $X$ and
$L_0$ alone is baseline-fixed. We drop $j$ likewise.

\begin{table*}[tp]
\centering
\caption{Diagnostics of artificial censoring in clone--censor--weight
analyses, grouped by the population that correct weights reconstruct.}
\label{tab:diag}
\footnotesize
\begin{tabularx}{\textwidth}{@{}L{3.4cm}YL{4.6cm}@{}}
\toprule
Diagnostic & Comparison in $\mathcal I_k$ & Zero target under a correct
censoring model \\
\midrule
\multicolumn{3}{@{}l}{\emph{Interval reconstruction (risk set before censoring)}} \\
Interval RRB, (\ref{eq:rrb}) & $\mathcal U$ weighted by $w/\hat\pi$ vs.\
$\mathcal R$ weighted by $w$ & Every history function (Proposition~1) \\
Censoring balance \cite{jackson2019} & $\mathcal C$ weighted by
$w/(1-\hat\pi)$ vs.\ $\mathcal U^c$ weighted by $w/\hat\pi$ & Every history
function if $\pi_k\le1-\varepsilon$ for the censorable individuals
(Proposition~1) \\
\addlinespace
\multicolumn{3}{@{}l}{\emph{Cumulative reconstruction (baseline population)}} \\
Cumulative RRB, (\ref{eq:cum}) & $\mathcal U\cup\mathcal Y$ weighted by
$\tilde W$ vs.\ the baseline population & Baseline-fixed functions without
loss to follow-up or weight truncation (Proposition~2) \\
Between-strategy balance, baseline covariates, (\ref{eq:diff}) &
$\mathcal U_g$ vs.\ $\mathcal U_{g'}$, both weighted by $W_k$ & Before
events; afterwards the depletion terms of (\ref{eq:deplete}) \\
End-of-grace SMD & As above, in the first interval after the grace period
\cite{maringe2020,buranupakorn2021} & As above \\
\addlinespace
\multicolumn{3}{@{}l}{\emph{No reconstruction target}} \\
Between-strategy balance, post-baseline covariates & As above, for $L_k$ &
None; differences include effects of the strategies \\
\bottomrule
\end{tabularx}

\medskip
\raggedright
$\mathcal R$, $\mathcal C$, $\mathcal U$: individuals at risk, censored and
retained in $\mathcal I_k$; $\mathcal U^c$: censorable retained
individuals; $\mathcal Y$: individuals with the outcome in an earlier
interval; $w=q_kW_{k-1}$: reference weight; $\tilde W$: weights carried
forward after the outcome (Section~\ref{sec:strategy}). RRB, risk-set
reconstruction balance; SMD, standardised mean difference.
\end{table*}

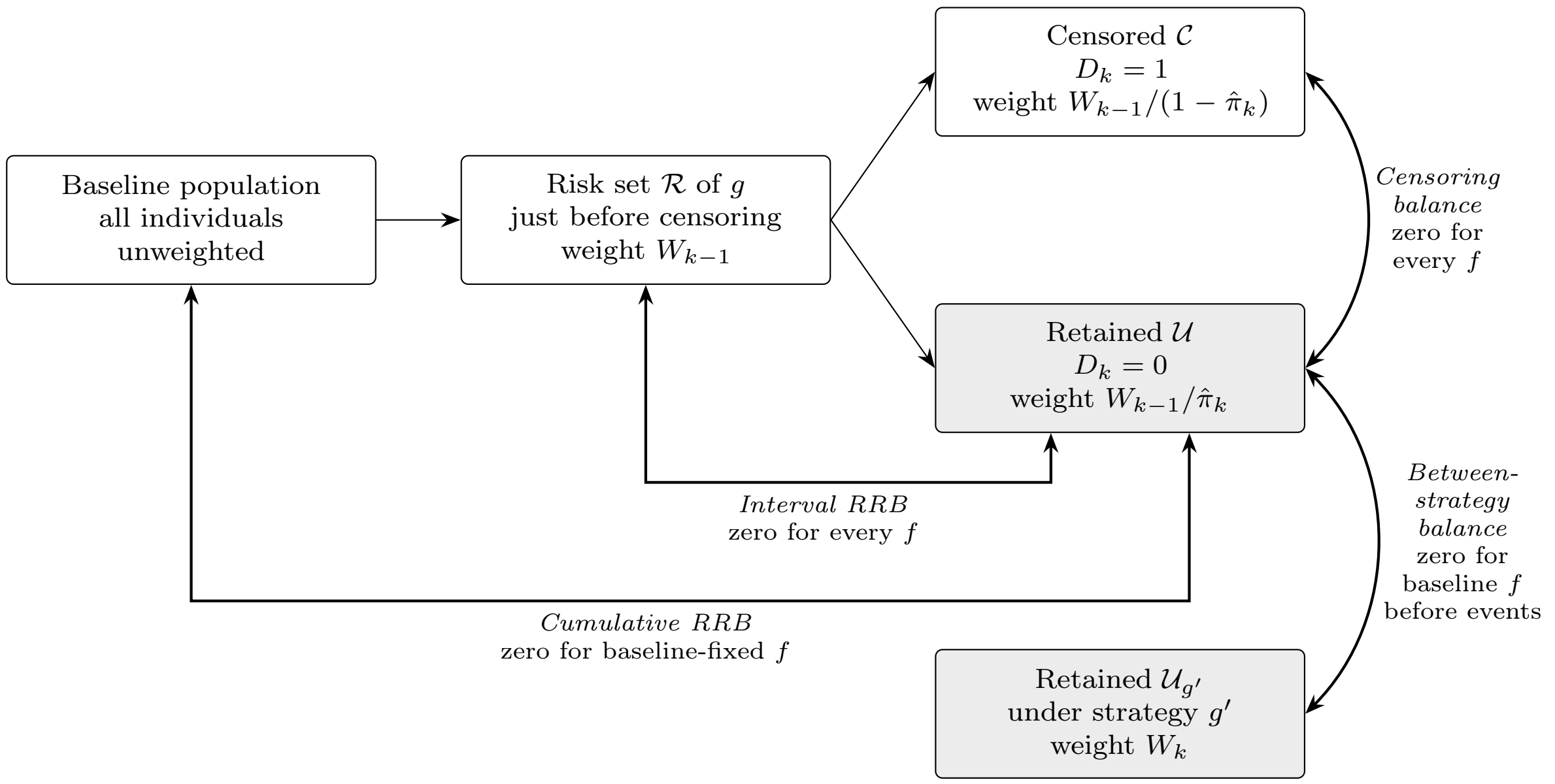
\begin{figure*}[tp]
\centering
\begin{tikzpicture}[
  box/.style={draw, rounded corners=2pt, align=center, font=\footnotesize,
    minimum height=1.3cm, text width=3.5cm, fill=white},
  cmp/.style={<->, >={Stealth[length=2mm]}, thick},
  flow/.style={->, >={Stealth[length=2mm]}},
  lab/.style={font=\scriptsize, align=center, inner sep=1.5pt}]
\node[box] (base) at (0, 0) {Baseline population\\all individuals\\unweighted};
\node[box] (pre) at (4.6, 0) {Risk set $\mathcal R$ of $g$\\just before censoring\\weight $W_{k-1}$};
\node[box] (cen) at (9.4, 1.5) {Censored $\mathcal C$\\\mbox{$D_k=1$}\\weight $W_{k-1}/(1-\hat\pi_k)$};
\node[box, fill=gray!15] (ret) at (9.4, -1.5) {Retained $\mathcal U$\\\mbox{$D_k=0$}\\weight $W_{k-1}/\hat\pi_k$};
\node[box, fill=gray!15] (oth) at (9.4, -5.0) {Retained $\mathcal U_{g'}$\\under strategy $g'$\\weight $W_k$};
\draw[flow] (base) -- (pre);
\draw[flow] (pre.east) -- (cen.west);
\draw[flow] (pre.east) -- (ret.west);
\draw[cmp] (pre.south) -- ++(0,-2.0) -| ($(ret.south)+(-0.7,0)$);
\node[lab, below=2pt] at ($(pre.south)+(1.8,-2.0)$) {\textit{Interval RRB}\\zero for every $f$};
\draw[cmp] (base.south) -- ++(0,-3.2) -| ($(ret.south)+(0.7,0)$);
\node[lab, below=2pt] at ($(base.south)+(4.6,-3.2)$) {\textit{Cumulative RRB}\\zero for baseline-fixed $f$};
\draw[cmp] (cen.east) to[bend left=45] node[lab, right] {\textit{Censoring}\\\textit{balance}\\zero for\\every $f$}
(ret.east);
\draw[cmp] (ret.east) to[bend left=45] node[lab, right]
{\textit{Between-}\\\textit{strategy}\\\textit{balance}\\zero for\\baseline $f$\\before events}
(oth.east);
\end{tikzpicture}
\caption{Populations compared by the diagnostics in $\mathcal I_k$ for
strategy $g$. Each box gives a population, its definition and its weight;
each double-headed arrow gives a diagnostic and the history functions $f$
for which it has a zero target under a correct censoring model. Shaded
boxes contain weighted retained individuals. Cumulative RRB also includes
the individuals with an earlier outcome, with carried-forward weights
(Section~\ref{sec:strategy}), and censoring balance compares the censored
individuals with the censorable retained individuals. The stabilising
numerator $q_k$ cancels within a stratum and is omitted. RRB, risk-set
reconstruction balance.}
\label{fig:schema}
\end{figure*}

\section{Reconstruction targets}
\label{sec:targets}
Table~\ref{tab:diag} and Figure~\ref{fig:schema} summarise the diagnostics
and their targets. Appendix S1 gives the derivations of (\ref{eq:resid}) and
(\ref{eq:deplete}).

\subsection{Interval reconstruction and censoring balance}
\label{sec:interval}
For $\mathcal I_k$, let $w_i=q_kW_{i,k-1}$ denote the reference weight, $\hat\pi_i$ the fitted
probability of remaining, and write $f_i=f_{ijk}$. The interval target is
that the retained individuals, reweighted by $1/\hat\pi_i$, reproduce the
risk set weighted by $w$ just before censoring. This local target checks the
current censoring step given the earlier weights, whether or not these are
correct. Define
\begin{gather*}
\mu_{\mathcal R}=\frac{\sum_{\mathcal R} w_if_i}{\sum_{\mathcal R} w_i},\qquad
\mu_{\mathcal U}=\frac{\sum_{\mathcal U} w_if_i\hat\pi_i^{-1}}
{\sum_{\mathcal U} w_i\hat\pi_i^{-1}},\\
\mu_{\mathcal C}=\frac{\sum_{\mathcal C} w_if_i(1-\hat\pi_i)^{-1}}
{\sum_{\mathcal C} w_i(1-\hat\pi_i)^{-1}},
\end{gather*}
and $s_{\mathcal R}$ the $w$-weighted standard deviation of the function in
the risk set. The stabilising numerator $q_k$ cancels out of the three
means and $s_{\mathcal R}$. Censoring balance (CB) is Jackson's
measure \cite{jackson2019}, the difference between the weighted means of
the censored and retained individuals, standardised here by
$s_{\mathcal R}$ so that RRB and CB share a denominator. Individuals who cannot be
censored ($\pi_k=1$), such as those who have already initiated under an
initiation strategy, never enter $\mathcal C$, so the weighted censored
individuals reproduce only the part of the risk set in which artificial
censoring is possible. Censoring balance therefore compares them with the
retained individuals of this part (censorable individuals), $\mathcal U^c$,
and $\mu_{\mathcal U^c}$ denotes $\mu_{\mathcal U}$ computed over
$\mathcal U^c$. Risk-set reconstruction
balance and censoring balance are
\begin{equation}
\mathrm{RRB}(k)=\frac{\mu_{\mathcal U}-\mu_{\mathcal R}}
{s_{\mathcal R}},\qquad
\mathrm{CB}(k)=\frac{\mu_{\mathcal C}-\mu_{\mathcal U^c}}
{s_{\mathcal R}}.
\label{eq:rrb}
\end{equation}

\emph{Proposition 1 (interval target).} Suppose that $\hat\pi$ converges in
probability, uniformly, to $\pi_k(H_k)$ with
$\pi_k\ge\varepsilon>0$, and that $w$ converges uniformly to a bounded function of $H_k$. Then
$\mathrm{RRB}(k)\to_p0$ for every $f(H_k)$, including functions of post-baseline covariates, with
finite mean and finite, positive weighted
variance in the risk set. If, in addition,
$\pi_k\le1-\varepsilon$ for the censorable individuals, then $\mathrm{CB}(k)\to_p0$
(proof in Appendix S1).

With the residual
$e_i=(1-D_i)-\hat\pi_i$, exactly,
\begin{equation}
\begin{gathered}
\sum_{\mathcal U}\frac{w_if_i}{\hat\pi_i}-\sum_{\mathcal R}w_if_i
=\sum_{\mathcal R}\frac{w_if_ie_i}{\hat\pi_i},\\
\mu_{\mathcal U}-\mu_{\mathcal R}=\frac{\sum_{\mathcal
R}w_i\hat\pi_i^{-1}(f_i-\mu_{\mathcal R})e_i}
{\sum_{\mathcal U}w_i\hat\pi_i^{-1}},
\end{gathered}
\label{eq:resid}
\end{equation}
so the numerator of RRB is a $w/\hat\pi$-weighted covariance between the
function and the residual, each term of which has conditional mean zero under a correct
model.

The score equations of the logistic model, fitted without weights, give
$\sum f_ie_i=0$ over the rows of the fit for every modelled function. For
a history function outside the model, $\sum f_ie_i$ is the score for adding
$f$ to the censoring model. If the reference weights are constant in the
stratum and a model with an intercept is fitted to the stratum alone, as
when artificial censoring under a strategy occurs in a single interval, the
score equals $\sum_{\mathcal R}(f_i-\mu_{\mathcal R})e_i$, and the numerator
of RRB weights each term of this sum by $w_i/\hat\pi_i$, the weight of a
retained individual in the estimator. RRB of a modelled function therefore
departs from zero only through this weighting and through pooling over
intervals, and diagnostic information lies mainly in history functions
outside the model. For a modelled function, RRB in a single interval of a
pooled fit can still reveal an association with censoring that changes
across intervals. Weighting by $w_i/\hat\pi_i$ puts RRB on the scale of the
weighted risk-set mean that the estimator reconstructs. The unweighted score test,
which asks whether $f$ improves the fit of the censoring model, remains a
complementary check.

Suppose that $\hat\pi$ is constant within the stratum. Every individual
in the stratum is then censorable, so that $\mathcal U^c=\mathcal U$. The
means $\mu_{\mathcal U}$ and $\mu_{\mathcal C}$ reduce to the $w$-weighted
means of the retained and censored individuals, $\mu_{\mathcal R}$ is their
weighted average, and
\[
\mathrm{RRB}=-\frac{\sum_{\mathcal C}w_i}{\sum_{\mathcal R}w_i}\,\mathrm{CB}.
\]
When $\hat\pi$ varies, RRB and CB are different statistics with the same
zero target. We take interval RRB as the primary interval diagnostic. RRB requires
only $\pi_k\ge\varepsilon>0$, whereas CB
weights the censored individuals by $(1-\hat\pi_i)^{-1}$, which requires
$\pi_k\le1-\varepsilon$ for the censorable individuals and is large and
variable when censoring is rare. As the two statistics have different
scales, they should not share a threshold.

Truncation caps $W_{ik}$, which then differs from $w_i/\hat\pi_i$, and
interval RRB computed with the truncated weights has no zero target. It measures the imbalance that truncation
leaves. The model-weight version (\ref{eq:rrb}), which combines the truncated reference
weights $w$ with the untruncated inverse probabilities $1/\hat\pi$ of the
current interval, keeps the zero target of Proposition~1, as does censoring
balance.

\subsection{Cumulative reconstruction and between-strategy balance}
\label{sec:strategy}
The retained individuals $\mathcal U$ of $\mathcal I_k$ carry $W_k$ and
include those who are lost to follow-up or have the outcome later in the
interval. Let $\mathcal Y$ be the individuals who had the outcome while
retained in an earlier interval ($\tau_i<k$). A survivor in
$\mathcal I_k$ carries $Q_kW^u_{ik}$, an
individual with an earlier event only $Q_{\tau_i}W^u_{i\tau_i}$, and
pooling the two does not reproduce the baseline population when $Q$
changes over time, as it does sharply at a grace-period end. We therefore
carry individuals with an earlier event forward with
$\tilde W_{ik}=W_{i\tau_i}Q_k/Q_{\tau_i}=Q_kW^u_{i\tau_i}$ and set
$\tilde W_{ik}=W_{ik}$ on $\mathcal U$. A baseline-fixed function $f$ has unweighted mean $\mu^0$
and standard deviation $s^0$ in the baseline population, which is the
same for every strategy. Cumulative reconstruction balance compares
$\mu^0$ with the cumulative mean $\mu\cum(k)$, the $\tilde W$-weighted mean
of $f$ over $\mathcal U\cup\mathcal Y$,
\begin{equation}
\begin{gathered}
\mathrm{RRB}\cum(k)=\frac{\mu\cum(k)-\mu^0}{s^0},\\
\mu\cum(k)=\frac{\sum_{\mathcal U\cup\mathcal Y}
\tilde W_{ik} f_i}{\sum_{\mathcal U\cup\mathcal Y}\tilde W_{ik}}.
\end{gathered}
\label{eq:cum}
\end{equation}

\emph{Proposition 2 (cumulative target).} Under the conditions of
Proposition~1 at every interval up to $k$, without loss to follow-up or
weight truncation, and
with the weights of individuals with an earlier event rescaled as above,
$\mathrm{RRB}\cum(k)\to_p0$ for every baseline-fixed function (proof in Appendix S1).

Under weight truncation, cumulative RRB
computed with the weights of the estimator has no zero target and measures
the imbalance in baseline-fixed functions that truncation leaves.

Between-strategy balance compares the weighted individuals retained under
two strategies $g$ and $g'$,
\[
B_{gg'}(k)=\frac{\mu^W_{g}(k)-\mu^W_{g'}(k)}{s_{gg'}(k)},\qquad
\mu^W_{g}(k)=\frac{\sum_{\mathcal
U_g}W_{ik}f_i}{\sum_{\mathcal U_g}W_{ik}},
\]
with $s_{gg'}$ the pooled standard deviation, the square root of the
average of the two
$W_k$-weighted variances. If no individual has had the outcome while at
risk under either strategy before $\mathcal I_k$, then
$\mu^W_g(k)=\mu\cum_g(k)$ for both, and for a baseline-fixed
function
\begin{equation}
B_{gg'}(k)=\frac{s^0}{s_{gg'}(k)}\left\{\mathrm{RRB}\cum_g(k)-
\mathrm{RRB}\cum_{g'}(k)\right\},
\label{eq:diff}
\end{equation}
as $\mu^0$ cancels. Between-strategy balance of baseline covariates is
therefore a between-strategy difference of cumulative reconstructions and
inherits the zero target; the end-of-grace SMD used in practice
\cite{maringe2020,buranupakorn2021} is (\ref{eq:diff}) at one time point.
Once individuals at risk have had the outcome, $\mathcal U_g$
represents only the survivors of the reweighted population. Let
$\tilde F_g(k)$ be the $\tilde W$-weighted share of individuals with an
earlier event in (\ref{eq:cum}), and $\mu^Y_g(k)$ their $\tilde W$-weighted
mean of $f$. As
$\mu\cum_g=\tilde F_g\mu^Y_g+(1-\tilde F_g)\mu^W_g$
holds exactly, (\ref{eq:diff}) extends to
\begin{equation}
\begin{aligned}
B_{gg'}(k)&=\frac{s^0}{s_{gg'}(k)}\bigl\{\mathrm{RRB}\cum_g(k)
-\mathrm{RRB}\cum_{g'}(k)\\
&\qquad{}+\Delta_g(k)-\Delta_{g'}(k)\bigr\},\\
\Delta_g(k)&=\frac{\mu^W_g-\mu\cum_g}{s^0}\\
&=-\frac{\tilde F_g}{1-\tilde F_g}\,
\frac{\mu^Y_g-\mu\cum_g}{s^0}.
\end{aligned}
\label{eq:deplete}
\end{equation}
We call $\Delta_g(k)$ the depletion term of strategy $g$.

\emph{Remark (depletion term).} Under the conditions of Proposition~2,
$\Delta_g(k)$ converges to
$-\{F_g/(1-F_g)\}\{\bar\mu^Y_g-E(f)\}/\mathrm{SD}(f)$, where
$F_g$ is the weighted cumulative incidence before $\mathcal I_k$ under
strategy $g$ and $\bar\mu^Y_g$ the limit of the mean of $f$ among these
events (Appendix S1). The limit is non-zero when events have occurred and
$f$ is prognostic.

Under a correct censoring model, the end-of-grace SMD can therefore be
non-zero once events have occurred, unless the depletion terms of the two
strategies cancel. The term formalises the depletion of susceptibles, which makes even
randomised groups differ in prognostic factors among survivors
\cite{hernan2010,aalen2015}, and the exclusion of individuals with an event
that the primer notes \cite{gaber2024}.

\emph{Remark (loss to follow-up).} Under the conditions of
Proposition~2 except that individuals can be lost to follow-up, cumulative
RRB converges to the limit of $-\lambda/(1-\lambda)$ times the standardised
difference between the lost individuals and the baseline population, where
$\lambda$ is the weighted share of the individuals lost while retained
before $\mathcal I_k$, carried forward as those with an earlier event
(Appendix S1).

The difference is
non-zero even under completely random loss once prognostic individuals
have had the outcome, as the lost individuals resemble the survivors at the
times of loss. Weighting by the inverse probabilities of remaining in
follow-up restores the target \cite{hernan2004,howe2016}.

Because between-strategy balance is a difference, it also cancels
deviations common to both strategies, such as a shared misspecification of
the censoring model; cumulative RRB compares each strategy with the fixed
baseline population and contains neither the depletion term nor this
cancellation. For a post-baseline covariate $L_k$, $B_{gg'}$
contains effects of the strategies on $L_k$ even under perfect weighting
\cite{rosenbaum1984,robins1986,daniel2013} and is descriptive, without a
zero target.

\subsection{Unmeasured variables}
\label{sec:unmeasured}
\emph{Proposition 3 (unmeasured variables).} Suppose that artificial censoring
depends on an unmeasured $U$ that also affects the outcome, and that $\hat\pi$
converges to $\pi_k(H_k)$. Then the targets of Propositions~1
and~2 are zero, but the weighted risk is in general biased for the risk under
the strategy (proof in Appendix S1).

The diagnostics thus detect misspecification in the measured history, not
violations of exchangeability.

\section{A null reference for the interval diagnostics}
\label{sec:null}
Interval RRB and censoring balance are summarised for each strategy by
their maximum absolute value over the strata with artificial censoring,
for example $\mathrm{MaxIB}=\max_k|\mathrm{RRB}(k)|$; where stated, the
maximum is also taken over history functions. In a sample, the diagnostics
fluctuate around their zero targets, so these maxima are positive even
under a correct censoring model.

\emph{Remark (null SD).} Under a correct model, with $\pi_k$ roughly
constant in a stratum and a function unrelated to the weights, the
standard deviation (SD) of RRB is approximately $\{(1-\pi_k)/\mathrm{ESS}\}^{1/2}$ (Appendix S1).

The null SD thus grows with the share censored and with weight
concentration, and the maximum of $m$ approximately independent standard
normal values is about $\sqrt{2\log m}$. A threshold such as 0.1
therefore has no fixed false-positive rate. The
sampling variability of SMDs has been studied for two-group comparisons at
a single time point, with a Bonferroni-corrected critical value for the
maximum studentised excess of the SMDs over 0.1 \cite{hripcsak2025}, and
a randomisation-based omnibus test covers stratified and clustered designs
\cite{hansen2008}. In person-interval data by strategy, censoring is
sequential and the records of one individual enter several intervals and
strata, which makes a resampling reference natural.

The package function \texttt{rrb\_null()} provides a parametric reference
distribution for chance imbalance under a correct censoring model,
hereafter the null reference. In each stratum, conditional on the observed
risk set and reference weights $w$, it redraws the censoring indicators,
$D^*_k\sim\mathrm{Bernoulli}(1-\hat\pi)$, on every row where censoring is
possible, refits the censoring model and recomputes interval RRB, in its
model-weight version (\ref{eq:rrb}) under weight truncation, and censoring
balance, $n_{\mathrm{draws}}$ times. For both statistics, it
reports each stratum's null SD and $z$, the ratio of the observed value to
it, and, for each strategy, the studentised maximum, the largest $|z|$ over
its strata and the history functions (excluding strata with a null SD of
0), with the 95th percentile of the same maximum over the draws and the
exceedance probability; the same quantities are reported for MaxIB. The
records of one individual under different strategies are coupled through
one uniform variate per individual and interval, which keeps every row's
marginal $1-\hat\pi$ and is approximate when the strategy-specific models
disagree (Appendix S2). Redrawn censoring is not propagated to later risk sets
or to $W_{k-1}$, as propagation would require models for the covariates
and outcomes of individuals after their censoring, so the reference
applies to the interval target only.

Each strategy is judged against the reference of its own censoring model
by its studentised maximum. The null SD of RRB differs between strata,
through the share censored and the ESS (Remark above), and between history
functions, so MaxIB is dominated by the strata with much censoring, whereas
division by the null SD puts every stratum and history function on the
same footing. MaxIB
is reported as a descriptive summary on the scale of the diagnostic
(Appendix S2).

In an oracle setting (fixed risk sets and weights, known censoring
probabilities, no refit), the redraw gives the exact conditional null
distribution of each stratum's RRB. The implementation replaces the known
probabilities by fitted ones, refits the model and holds the realised
dependence between strata fixed, so its calibration is examined
empirically, with an oracle check and a per-strategy decomposition
(Table S1; Section~\ref{sec:cal}). The null reference is therefore a diagnostic
reference distribution with approximate calibration, and its exceedance
probability indicates how unusual the observed imbalance is under the
fitted model; it is not the size of a formal test. An
exceedance is evidence of misspecification of the censoring model with
respect to the measured history, subject to this calibration and to the
multiplicity of the strata and history functions examined. A
non-exceedance is weak evidence of a correct model, and its reassurance
depends on the same calibration, on the ESS and on the distribution of the
history function.

\section{Simulation study}
\label{sec:sim}
We follow the ADEMP structure \cite{morris2019}.

\subsection{Aims}
The simulation study has three parts. Part 1 shows that the
interval diagnostics of Section~\ref{sec:interval} respond to an omitted
term's effect on censoring, whether or not the term is prognostic, whereas
bias requires outcome-related selection. Part 2 examines the calibration of
the null reference of Section~\ref{sec:null}, with many history functions
and under unstable weights. Part 3 examines the cumulative and between-strategy
targets of Section~\ref{sec:strategy} under treatment$\to$covariate
effects, under depletion and under loss to follow-up that the weights
ignore, and the variability of cumulative RRB.

\subsection{Data-generating mechanisms}
A baseline covariate $X\sim N(0,1)$ and a
time-varying covariate $L_k$, autoregressive and affected by prior treatment
with coefficient $\alpha_2$, determine absorbing treatment initiation
(logistic in $L_k$, $L_k^2$, $X$, $L_kX$ and $L_{k-1}$, coefficients
$\beta$), optional covariate-dependent loss to follow-up, and a binary
outcome over 12 intervals (logistic, coefficients $\gamma$; Appendix S2),
in the order of Section~\ref{sec:notation}. The initiation hazard does not
depend on the interval, so the censoring models contain no time terms. The
strategies are initiation in intervals 0--2 ($g_1$), initiation in
intervals 0--8 ($g_2$), and no initiation through interval 8 ($g_3$). Under
the initiation strategies, an individual who has not initiated by the last
interval of the window deviates in that interval; under $g_3$, an individual
deviates at initiation in intervals 0--8, and initiation from interval 9
onwards is allowed. The contrast of $g_2$ against $g_1$ concerns the length
of the initiation window, and that of $g_3$ against $g_2$ initiation within
the window against no initiation.

\begin{table*}[tp]
\centering
\caption{Simulation scenarios of Parts 1--3: omitted term with censoring
strength / outcome strength, parameter overrides of the defaults, analysis
artificial-censoring model, sample size, replicates and bootstrap resamples.}
\label{tab:grid}
\scriptsize
\setlength{\tabcolsep}{3pt}
\begin{tabularx}{\textwidth}{@{}llYYrrr@{}}
\toprule
Scenario & Strengths & Overrides & Censoring model & $n$ & Replicates & $n_{\mathrm{boot}}$ \\
\midrule
\multicolumn{7}{@{}l}{\emph{Part 1: censoring-model misspecification and estimation bias}} \\
Correct specification & -- & -- & $\mathrm{M}_0$ & 2{,}000 & 1{,}000 & 500 \\
Omitted $L_k$ & weak / none & $\beta_1=0.4,\ \beta_2=\beta_5=0,\ \gamma_1=0$ & $X$ & 2{,}000 & 200 & 100 \\
Omitted $L_k$ & weak / strong & $\beta_1=0.4,\ \beta_2=\beta_5=0,\ \gamma_1=1.5$ & $X$ & 2{,}000 & 200 & 100 \\
Omitted $L_k$ & strong / none & $\beta_1=1,\ \beta_2=\beta_5=0,\ \gamma_1=0$ & $X$ & 2{,}000 & 200 & 100 \\
Omitted $L_k$ & strong / strong & $\beta_1=1,\ \beta_2=\beta_5=0,\ \gamma_1=1.5$ & $X$ & 2{,}000 & 200 & 100 \\
Omitted $L_k^2$ & weak / none & $\beta_5=0,\ \beta_2=0.3,\ \gamma_0=-4.5,\ \gamma_4=0$ & $L_k + X$ & 2{,}000 & 200 & 100 \\
Omitted $L_k^2$ & weak / strong & $\beta_5=0,\ \beta_2=0.3,\ \gamma_0=-4.5,\ \gamma_4=1$ & $L_k + X$ & 2{,}000 & 200 & 100 \\
Omitted $L_k^2$ & strong / none & $\beta_5=0,\ \beta_2=0.6,\ \gamma_0=-4.5,\ \gamma_4=0$ & $L_k + X$ & 2{,}000 & 200 & 100 \\
Omitted $L_k^2$ & strong / strong & $\beta_5=0,\ \beta_2=0.6,\ \gamma_0=-4.5,\ \gamma_4=1$ & $L_k + X$ & 2{,}000 & 200 & 100 \\
Omitted $L_{k-1}$ & weak / none & $\beta_6=0.6,\ \gamma_6=0$ & $\mathrm{M}_0$ & 2{,}000 & 200 & 100 \\
Omitted $L_{k-1}$ & weak / strong & $\beta_6=0.6,\ \gamma_6=1.2$ & $\mathrm{M}_0$ & 2{,}000 & 200 & 100 \\
Omitted $L_{k-1}$ & strong / none & $\beta_6=1.2,\ \gamma_6=0$ & $\mathrm{M}_0$ & 2{,}000 & 200 & 100 \\
Omitted $L_{k-1}$ & strong / strong & $\beta_6=1.2,\ \gamma_6=1.2$ & $\mathrm{M}_0$ & 2{,}000 & 200 & 100 \\
\addlinespace
\multicolumn{7}{@{}l}{\emph{Part 2: null-reference calibration and weight instability (with the correctly specified scenario of Part 1)}} \\
Correct specification & -- & -- & $\mathrm{M}_0$ & 300 & 1{,}000 & 500 \\
Correct specification & -- & -- & $\mathrm{M}_0$ & 8{,}000 & 500 & 200 \\
Near-violation of positivity & -- & $\beta_0=-2,\ \beta_1=1.2,\ \beta_2=0.4,\ \beta_3=0.6$ & $\mathrm{M}_0$ & 2{,}000 & 200 & 500 \\
Weight truncation, 95th percentile & -- & -- & $\mathrm{M}_0$ & 2{,}000 & 200 & 500 \\
Many correlated covariates & -- & $X_2,\dots,X_9$, $\rho=0.5$ & $\mathrm{M}_0 + X_2 + \dots + X_9$ & 300 & 200 & 100 \\
Extreme weights & -- & $\beta_0=-2,\ \beta_1=2,\ \beta_2=0.5,\ \beta_3=1$ & $\mathrm{M}_0$ & 2{,}000 & 200 & 100 \\
\addlinespace
\multicolumn{7}{@{}l}{\emph{Part 3: reconstruction targets}} \\
No treatment$\to$covariate effect & -- & $\alpha_2=0$ & $\mathrm{M}_0$ & 2{,}000 & 200 & 100 \\
Strong treatment$\to$covariate effect & -- & $\alpha_2=-1.5$ & $\mathrm{M}_0$ & 2{,}000 & 200 & 100 \\
Unweighted loss to follow-up & -- & $\delta_0=-2,\ \delta_1=1.5,\ \gamma_1=1$ & $\mathrm{M}_0$ & 2{,}000 & 200 & 100 \\
Many early events & -- & $\gamma_0=-2,\ \gamma_2=-3.5,\ \gamma_3=2.5$ & $\mathrm{M}_0$ & 2{,}000 & 200 & 500 \\
\bottomrule
\end{tabularx}

\medskip
\raggedright
Strengths: censoring strength / outcome strength, the coefficient of the
omitted term in the treatment (artificial-censoring) model / in the outcome
model. Censoring model: terms of the pooled logistic
model for remaining uncensored; $\mathrm{M}_0 = L_k + L_k^2 + X + L_kX$ is
correct for the base mechanism. Parameters ($\beta$, treatment; $\gamma$,
outcome; $\alpha$, covariate; $\delta$, loss to follow-up) and defaults:
Appendix S2. $n_{\mathrm{boot}}$: bootstrap resamples for the standard
errors of $\hat\theta_{21}$ and $\hat\theta_{32}$ and the percentile
intervals of cumulative RRB. Weight truncation: at the
95th percentile of the stabilised cumulative weights of retained
individuals, pooled over strategies and intervals. $X_2,\dots,X_9$: extra
baseline covariates with correlation $\rho$ with $X$ (Appendix S2).
\end{table*}

The three parts (Table~\ref{tab:grid}) have $n=2{,}000$
unless stated. Part 1 (censoring-model misspecification
and estimation bias) omits $L_k$, $L_k^2$ or $L_{k-1}$ from the
artificial-censoring model, one term at a time, and crosses the censoring
strength, the coefficient of the omitted term in the treatment model, which
determines censoring (weak or strong), with the outcome strength, its
coefficient in the outcome model (none or strong). For omitted $L_k$, the treatment model also
lacks $L_k^2$ and $L_kX$, and for omitted $L_k^2$ it lacks $L_kX$, so each
omitted term defines its own mechanism. With the correct model, Part 1 has
13 scenarios.

Part 2 (null-reference calibration and weight instability) has scenarios
with correct specification at $n=300$, 2,000 (shared with Part 1) and 8,000
(1,000, 1,000 and 500 replicates; 200 in every other scenario),
near-violation of positivity, weight
truncation and extreme weights, and one scenario at $n=300$ that adds eight
baseline covariates $X_2,\dots,X_9$ with correlation $\rho=0.5$ with $X$,
all modelled and monitored (16 history functions; about 175 combinations
of stratum and history function, against 87 in the base mechanism).

Part 3 (reconstruction targets) covers scenarios without and with a
strong treatment$\to$\allowbreak covariate effect, covariate-dependent loss
to follow-up that the weights ignore, and many early events under a correct model, which
make the depletion term large. Coefficients were chosen so that the predicted phenomena are visible.

\subsection{Estimands}
The true risk under a grace-period strategy ($g_1$, $g_2$)
is that of the natural grace-period regime, in which initiation occurs at
its natural time and is forced in the last interval of the grace period
for individuals who have not initiated \cite{wanis2024}; CCW targets this
regime for a strategy of initiation within a window
\cite{websterclark2025}. The true risk under $g_3$ is that of the regime
in which no individual initiates in intervals 0--8 and initiation follows
its natural course from interval 9 onwards. The estimands are the risk
differences at the end
of follow-up without loss to follow-up, $\theta_{21}$ ($g_2$ minus $g_1$)
and $\theta_{32}$ ($g_3$ minus $g_2$), whose true values were computed from
$5\times10^5$ individuals simulated under each regime (Monte Carlo
standard error (MCSE) about $7\times10^{-4}$).

\subsection{Methods}
Pooled logistic artificial-censoring models were fitted by
strategy over its censoring intervals. The correct model contained $L_k$, $L_k^2$, $X$ and $L_kX$ (and
$X_2,\dots,X_9$ in the scenario with many covariates). Risks were estimated
by a weighted discrete-time Kaplan--Meier estimator, with
normal-approximation 95\% confidence intervals (CIs) from a bootstrap standard error (SE)
\cite{austin2016,maringe2020} that resampled individuals, with their
records under all strategies, and refitted the censoring model. The
diagnostics, summarised as in Section~\ref{sec:null} over the intervals
with artificial censoring of each strategy, were interval RRB and
censoring balance (censorable individuals) for $X$, $L_k$, $L_k^2$,
$L_kL_{k-1}$, $L_k-L_{k-1}$, the cumulative mean of $L$, $I(L_k>0)$ and
$L_kX$ (and $X_2,\dots,X_9$); cumulative RRB of $X$ and $L_0$, summarised
by its maximum absolute value over the same intervals, the strategies and
the two functions; and the
end-of-grace SMD of $X$ at interval 3 for $g_2$ against $g_1$ and at
interval 9 for $g_3$ against $g_2$, the first intervals after the grace
periods of $g_1$ and $g_2$. In every replicate, the null reference (500
draws) determined, for each
strategy, whether the studentised maximum of interval RRB over all history
functions, that of each history function, that of censoring balance and
MaxIB exceeded its null 95th percentile. Under weight truncation, interval
RRB was computed with the truncated weights of the estimator, and the null
reference computed interval RRB and censoring balance with untruncated
inverse probabilities $1/\hat\pi$, as in the model-weight version (\ref{eq:rrb}). The bootstrap also gave a percentile
95\% interval for cumulative RRB of $X$ under each strategy at intervals 3
and 9. The bootstrap had 100 resamples, and 500 in the scenarios with correct
specification at $n=300$ and 2,000, near-violation of positivity, weight
truncation and many early events (200 at $n=8{,}000$). We recorded the smallest ESS/$|\mathcal R|$
and the largest W99 over all intervals and strategies, and the maximum
between-strategy SMD of $L$.

\subsection{Performance measures}
The performance measures for $\hat\theta_{21}$ and $\hat\theta_{32}$ were
the bias, the Monte Carlo standard deviation of the estimates (MCSD), the
absolute bias over the MCSD, the coverage and the ratio of the mean
bootstrap SE to the MCSD. Each diagnostic was summarised by its mean.
Biases, coverages and means are reported with their MCSEs. For the
factorial design of Part 1, the main effects and interaction of censoring
strength and outcome strength were computed for the bias and for interval RRB. The exceedance rate is the
proportion of replicates in which a maximum exceeded its null 95th
percentile or, for cumulative RRB, the bootstrap interval excluded 0. It is
a false-positive rate (nominal 0.05) under a correct model and a detection
rate otherwise. Replicates in which the studentised maximum was undefined,
because every null draw reproduced the observed RRB in every stratum, were
excluded. As each omitted term defines its own mechanism, mean diagnostics
are compared only among the scenarios with the same omitted term.

\begin{figure*}[tp]
\centering
\includegraphics[width=\textwidth]{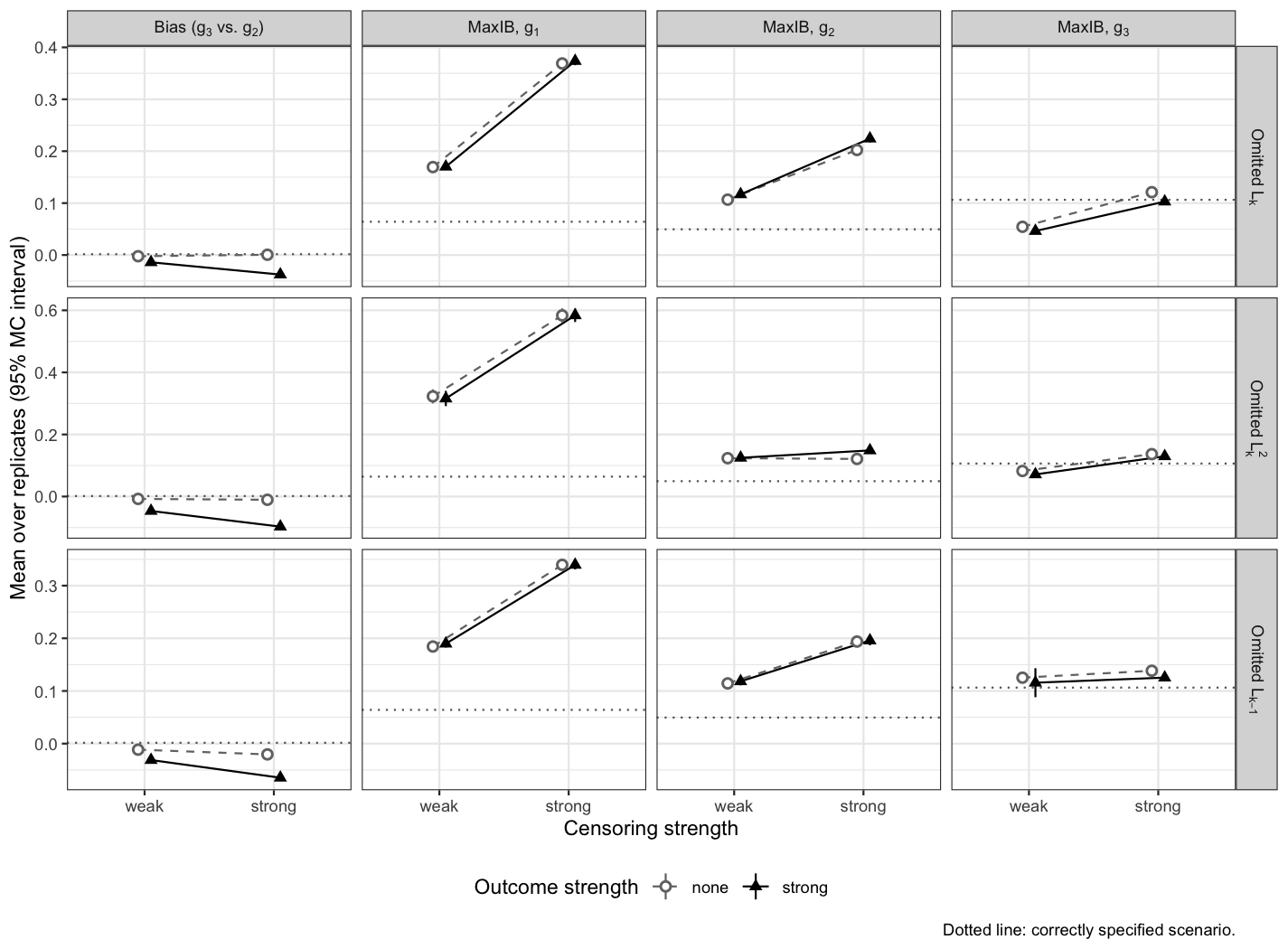}
\caption{Misspecification factorial (Part 1, $n=2{,}000$, 200 replicates per
scenario). Rows, omitted term of the artificial-censoring model; columns,
bias of $\hat\theta_{32}$ ($g_3$ vs.\ $g_2$) and MaxIB of interval RRB of
$g_1$, $g_2$ and $g_3$, each on its own scale. Horizontal axis, censoring
strength; symbol, line type and grey level, outcome strength (Section~\ref{sec:sim}).
Points, means over replicates with 95\% Monte Carlo intervals; dotted
lines, the correctly specified base scenario (1,000 replicates), a visual
guide only, because each misspecified scenario changes coefficients of the
base mechanism (Table~\ref{tab:grid}); MaxIB is comparable within a row.}
\label{fig:fact}
\end{figure*}

\begin{table*}[tp]
\centering
\caption{Part 1, bias and coverage of $\hat\theta_{21}$ and $\hat\theta_{32}$ and
detection rate of each strategy.}
\label{tab:simres}
\small
\resizebox{\textwidth}{!}{\begin{tabular}{@{}lllrrrrrrr@{}}
\toprule
 & & & \multicolumn{2}{c}{$g_2$ vs.\ $g_1$ ($\hat\theta_{21}$)} & \multicolumn{2}{c}{$g_3$ vs.\ $g_2$ ($\hat\theta_{32}$)} & \multicolumn{3}{c}{Detection rate} \\
\cmidrule(lr){4-5}\cmidrule(lr){6-7}\cmidrule(l){8-10}
Omitted term & \hd{Censoring\\strength} & \hd{Outcome\\strength} & \hd{$|$Bias$|$/\\MCSD} & Coverage & \hd{$|$Bias$|$/\\MCSD} & Coverage & $g_1$ & $g_2$ & $g_3$ \\
\midrule
Correct specification & -- & -- & 0.02 & 0.945 & 0.06 & 0.937 & 0.022 & 0.018 & 0.048 \\
\addlinespace
Omitted $L_k$ & weak & none & 0.10 & 0.960 & 0.15 & 0.955 & 0.540 & 0.275 & 0.845 \\
 & weak & strong & 0.19 & 0.955 & 0.90 & 0.855 & 0.540 & 0.205 & 0.875 \\
 & strong & none & 0.08 & 0.940 & 0.03 & 0.955 & 0.995 & 0.805 & 1.000 \\
 & strong & strong & 0.49 & 0.925 & 1.70 & 0.485 & 1.000 & 0.820 & 1.000 \\
\addlinespace
Omitted $L_k^2$ & weak & none & 0.20 & 0.915 & 0.55 & 0.880 & 0.880 & 0.515 & 1.000 \\
 & weak & strong & 0.59 & 0.895 & 1.72 & 0.625 & 0.785 & 0.370 & 0.990 \\
 & strong & none & 0.26 & 0.905 & 0.82 & 0.850 & 1.000 & 0.985 & 1.000 \\
 & strong & strong & 1.36 & 0.740 & 3.87 & 0.020 & 1.000 & 0.930 & 1.000 \\
\addlinespace
Omitted $L_{k-1}$ & weak & none & 0.06 & 0.945 & 0.51 & 0.860 & 0.605 & 0.195 & 0.770 \\
 & weak & strong & 0.05 & 0.925 & 1.27 & 0.600 & 0.675 & 0.185 & 0.835 \\
 & strong & none & 0.01 & 0.940 & 1.00 & 0.730 & 0.990 & 0.505 & 1.000 \\
 & strong & strong & 0.10 & 0.950 & 3.05 & 0.155 & 0.980 & 0.435 & 1.000 \\
\bottomrule
\end{tabular}}

\medskip
\raggedright
Coverage: of the 95\% CI (MCSE at most 0.035). Detection rate: proportion
of replicates in which the studentised maximum of interval RRB exceeded its
null 95th percentile (binomial SE at most 0.035); under correct
specification, it is a false-positive rate with nominal level 0.05. CI, confidence interval; MCSD, Monte Carlo standard
deviation; RRB, risk-set reconstruction balance.
\end{table*}

\section{Simulation results}
\label{sec:simres}
\subsection{Part 1: censoring-model misspecification and estimation bias}
\label{sec:fact}
Substantial bias arose only when the omitted term affected both censoring and
outcome, directly or through correlated prognostic terms, and it was
concentrated in $\hat\theta_{32}$ (Figure~\ref{fig:fact},
Table~\ref{tab:simres}). The interaction contrast of its bias was
$-0.048$ to $-0.025$ (MCSE 0.003; Table S5). With strong
censoring and outcome strengths, its bias was 1.7 to 3.9 MCSDs and its coverage 0.020 to
0.485. With an outcome strength of none, its absolute bias was at most
1.00 MCSD (omitted $L_{k-1}$ with strong censoring strength; coverage
0.730). Under $g_3$, a misspecified model acts on
the weights in each of the nine censoring intervals, which is consistent
with the larger bias of $\hat\theta_{32}$.

The interval diagnostics followed mainly the censoring strength. Over the three
omitted terms, the main effect of censoring strength on MaxIB
was 0.15 to 0.26 under $g_1$, 0.01 to 0.10 under $g_2$ and 0.01 to 0.06
under $g_3$, and the main effects of outcome strength were small ($-0.013$ to
0.016; Table S5).
Detection depended on the strategy and rose with censoring strength
(Table~\ref{tab:simres}). It was highest under $g_3$
(0.770 to 1.000), whose model is fitted over nine intervals, and lowest
under $g_2$ (0.185 to 0.985), whose single censoring interval has fewer
censorable individuals than that of $g_1$ (Table S1). Omitted $L_{k-1}$ was
detected mainly through $L_k-L_{k-1}$, whose own studentised maximum
exceeded its reference in 0.400 to 1.000 of replicates (Table S2), whereas
under $g_3$ MaxIB exceeded its own reference in only 0.075 to 0.130
(Table S2).
Censoring balance, with its own null reference, detected the same
mechanisms. Neither statistic was uniformly more sensitive (Table S4).

\begin{table*}[tp]
\centering
\caption{Part 2, calibration of the null reference and estimator
performance under correct censoring models.}
\label{tab:calres}
\small
\setlength{\tabcolsep}{4pt}
\resizebox{\textwidth}{!}{\begin{tabular}{@{}lrrrrrrrrrrrrr@{}}
\toprule
 & & \multicolumn{3}{c}{$g_2$ vs.\ $g_1$ ($\hat\theta_{21}$)} & \multicolumn{3}{c}{$g_3$ vs.\ $g_2$ ($\hat\theta_{32}$)} & \multicolumn{3}{c}{\hd{Exceedance\\rate, RRB}} & \multicolumn{3}{c}{\hd{Exceedance\\rate, CB}} \\
\cmidrule(lr){3-5}\cmidrule(lr){6-8}\cmidrule(lr){9-11}\cmidrule(l){12-14}
Scenario & $n$ & \hd{$|$Bias$|$/\\MCSD} & \hd{SE\\ratio} & Coverage & \hd{$|$Bias$|$/\\MCSD} & \hd{SE\\ratio} & Coverage & $g_1$ & $g_2$ & $g_3$ & $g_1$ & $g_2$ & $g_3$ \\
\midrule
Correct specification & 300 & 0.01 & 0.91 & 0.896 & 0.05 & 0.85 & 0.923 & 0.022 & 0.026 & 0.036 & 0.022 & 0.006 & 0.041 \\
Correct specification & 2{,}000 & 0.02 & 0.99 & 0.945 & 0.06 & 0.70 & 0.937 & 0.022 & 0.018 & 0.048 & 0.019 & 0.020 & 0.057 \\
Correct specification & 8{,}000 & 0.04 & 1.01 & 0.954 & 0.05 & 0.85 & 0.940 & 0.030 & 0.028 & 0.050 & 0.024 & 0.036 & 0.062 \\
Near-violation of positivity & 2{,}000 & 0.08 & 1.05 & 0.925 & 0.36 & 0.60 & 0.650 & 0.015 & 0.020 & 0.060 & 0.025 & 0.015 & 0.055 \\
Weight truncation, 95th percentile & 2{,}000 & 0.93 & 1.06 & 0.895 & 1.64 & 0.99 & 0.615 & 0.015 & 0.045 & 0.050 & 0.015 & 0.035 & 0.065 \\
\addlinespace
Many correlated covariates & 300 & 0.06 & 1.30 & 0.965 & 0.06 & 0.76 & 0.900 & 0.015 & 0.107 & 0.040 & 0.020 & 0.027 & 0.025 \\
Extreme weights & 2{,}000 & 0.10 & 0.81 & 0.910 & 0.58 & 0.62 & 0.500 & 0.040 & 0.035 & 0.040 & 0.030 & 0.010 & 0.030 \\
\bottomrule
\end{tabular}}

\medskip
\raggedright
SE ratio: mean bootstrap SE divided by the MCSD. Coverage: of the 95\% CI
(MCSE at most 0.035). Exceedance rate: as the detection rate of
Table~\ref{tab:simres} (binomial SE at most 0.023). Under $g_2$, replicates
with a null SD of 0 in every stratum are excluded (2 of 1,000 at $n=300$; 13
of 200 with many correlated covariates). CB, censoring balance over the
censorable individuals; CI, confidence interval; MCSD,
Monte Carlo standard deviation; MCSE, Monte Carlo standard error; RRB, risk-set
reconstruction balance; SD,
standard deviation; SE, standard error.
\end{table*}

\begin{figure*}[tp]
\centering
\includegraphics[width=\textwidth]{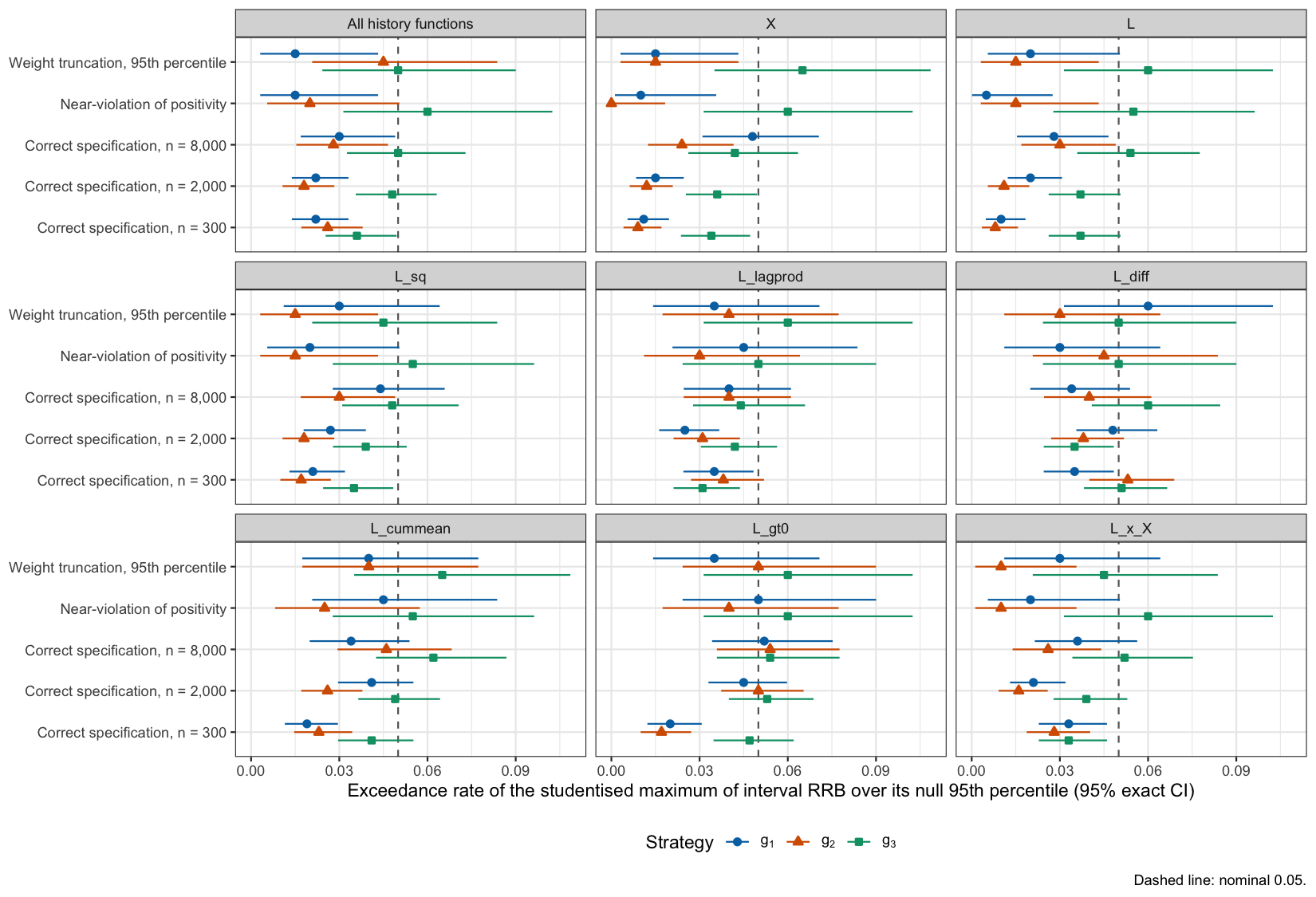}
\caption{Calibration of the null reference under correct censoring models
(Part 2; 1,000 replicates with correct specification at $n=300$ and
2,000, 500 at $n=8{,}000$, and 200 under near-violation of positivity and
under weight truncation). Rate at which the studentised maximum of
interval RRB of each strategy exceeded its null 95th percentile (500 draws;
Section~\ref{sec:null}), for all history functions (first panel) and for
each: $X$, $L_k$ (\texttt{L}), $L_k^2$
(\texttt{L\_sq}), $L_kL_{k-1}$ (\texttt{L\_lagprod}), $L_k-L_{k-1}$
(\texttt{L\_diff}), the cumulative mean of $L$ (\texttt{L\_cummean}),
$I(L_k>0)$ (\texttt{L\_gt0}) and $L_kX$ (\texttt{L\_x\_X}). Circles,
$g_1$; triangles, $g_2$; squares, $g_3$; exact binomial 95\% CIs; dashed
line, nominal 0.05. Weight truncation: untruncated inverse probabilities $1/\hat\pi$ of
the current interval.}
\label{fig:cal}
\end{figure*}

\subsection{Part 2: null-reference calibration and weight instability}
\label{sec:cal}
Under correct censoring models, the exceedance rate of the studentised
maximum was 0.035 to 0.060 under $g_3$ and 0.010 to 0.045 under $g_1$ and
$g_2$, and that of censoring balance 0.006 to 0.065 (Figure~\ref{fig:cal},
Table~\ref{tab:calres} and Table S2); the exception was $g_2$ with 16
history functions (below). For single history functions, the exceedance
rate was 0.000 to 0.085 (Table S2). MaxIB averaged 0.156 to 0.195 at $n=300$ and 0.362 for $g_3$ under
near-violation of positivity, above the conventional 0.1 (Table S3).

In the oracle setting of Section~\ref{sec:null} (scheme S5 of Table S1),
exceedance of the studentised maximum was 0.038 to 0.061 for the three
strategies. The per-strategy decomposition of Table S1 located the
shortfall of $g_1$ and $g_2$ in the redraw from the fitted probabilities.
At $n=2{,}000$, the redraw from the true probabilities with refitting gave
0.050 to 0.065, and the redraw from $\hat\pi$ without refitting gave 0.002
to 0.038, so refitting is necessary. Under $g_1$ and $g_2$, the shortfall
was largest for the modelled terms (0.011 to 0.027, against 0.025 to 0.050 for the
other functions; Table S2), consistent with the score equations nearly
balancing them (Section~\ref{sec:interval}).

With 16 history functions at $n=300$, the exceedance rate
under $g_2$ was 0.107 (SE 0.023), above the nominal 0.05, and at most
0.040 under the other strategies and for censoring balance
(Table~\ref{tab:calres}); the single history functions of $g_2$ exceeded
their references in 0.043 to 0.171 of replicates.

Under correct specification at $n=2{,}000$ and 8,000, coverage was 0.937
to 0.954. At $n=300$, it was 0.896 for $\hat\theta_{21}$ and 0.923 for
$\hat\theta_{32}$ (MCSE 0.010 and 0.008), below the nominal 0.95. The SE
ratio was 0.91 to 1.01 for $\hat\theta_{21}$ and 0.70 to 0.85 for
$\hat\theta_{32}$.
A few outlying replicates inflated the MCSD of $\hat\theta_{32}$; with the
interquartile range, its SE ratio was 1.03 to 1.11 (Appendix S2).
Truncation gave the most favourable weights (smallest ESS/$|\mathcal R|$ 0.191,
largest W99 1.55; base scenario 0.113 and 9.38) but an absolute bias of
0.93 MCSD for $\hat\theta_{21}$ (coverage 0.895) and of 1.64 MCSD for
$\hat\theta_{32}$ (coverage 0.615). With the weights of the
estimator, MaxIB of interval RRB was 0.274 and 0.201 under $g_1$ and $g_2$ (base
scenario 0.064 and 0.050), cumulative RRB 0.266 (0.095), and the two end-of-grace SMDs were 0.28
and 0.44 (0.02 and 0.09; Table S3). With the model weights
of the null reference, MaxIB was close to that of the base scenario under
$g_1$ and $g_2$ and smaller under $g_3$ (Table S3).
Under extreme weights (coverage of $\hat\theta_{32}$ 0.500; SE ratio 0.62)
and near-violation of positivity (0.650; 0.60), the exceedance rate was
0.015 to 0.060, so the null reference does not flag extreme weights.

\subsection{Part 3: reconstruction targets}
\label{sec:cumres}
Without a treatment$\to$covariate effect, the maximum between-strategy SMD
of $L$ for $g_3$ against $g_2$ was 0.11; it was 0.90 with the default
effect and 2.64 with the strong effect (0.11 to 1.27 for the other
contrast). MaxIB of interval RRB, the bias of both contrasts (at most 0.10
MCSD) and the exceedance rate (0.010 to 0.055) did not change
systematically (Tables S2 and S3). With $L_k$ omitted (Part 1), the same SMD was
0.82 to 0.88, so between-strategy balance of $L$ would have flagged the valid
analysis with a strong effect and missed the biased ones.

By the end of the grace period of $g_1$, 5.4\% of individuals had the
outcome in the base scenario and 35.8\% with many early events, under which
the depletion term became visible. The end-of-grace SMD of $X$ was 0.056 at
interval 3 and 0.403 at interval 9, against 0.015 and 0.089 in the base scenario, whereas
cumulative RRB was 0.049 (base scenario 0.095). The exceedance rate was
0.010 to 0.050 (Table S2), the bias was 0.23 MCSD for $\hat\theta_{21}$
and 0.05 MCSD for $\hat\theta_{32}$, and coverage was 0.925 and 0.945
(Table S3).

When the weights ignored covariate-dependent loss to follow-up, the bias of
$\hat\theta_{32}$ was 2.5 MCSDs (coverage 0.290) and that of
$\hat\theta_{21}$ 0.9 MCSD (coverage 0.890). MaxIB of interval RRB (at most 0.090;
exceedance rate 0.015 to 0.035; Tables S2 and S3) and the end-of-grace SMD
at interval 3 (0.028) stayed small because the artificial-censoring model
was correct. The end-of-grace SMD at interval 9 was 0.234, smaller than
under many early events with a correct model, so this SMD does not
separate loss to follow-up from depletion. Cumulative RRB was 0.448
(Table S3).

Under correct models the bootstrap interval of cumulative RRB of $X$
excluded 0 in 0.017 to 0.095 of replicates at interval 3, after the end of
the grace period of $g_1$, and in 0.016 to 0.211 at interval 9, after the
end of the grace period of $g_2$, highest under $g_3$ after prolonged
censoring with concentrated weights (Table S6). The individual bootstrap
interval is thus anti-conservative for the cumulative target.

\section{Case study}
\label{sec:emp}
\subsection{Setting}
\label{sec:empset}
We re-analysed the public cohort of the Webster-Clark CCW tutorial
\cite{websterclark2025}, which was built from the 2008--2010 Data
Entrepreneurs' Synthetic Public Use File (DE-SynPUF) of the Centers for
Medicare \& Medicaid Services. The cohort comprises 1,590 patients
discharged after myocardial infarction, followed for initiation of
antiplatelet treatment and for rehospitalisation or death within 180 days.
We linked the cohort file, which records baseline covariates only, to the
DE-SynPUF claims and derived, for each individual and interval, three
indicators of healthcare use in the 10 days before the interval began,
namely a physician claim, a hospital outpatient claim and a dispensing of
a drug other than the index antiplatelet (present in 54\%, 13\% and 58\% of
individual-intervals). From the dispensing history we also derived three
depth functions, namely the indicator of the preceding window (10 to 19
days before the interval; present in 54\%), the number of dispensings in
the 30 days before the interval (mean 3.1) and the number since discharge
(mean 8.9).

Three strategies (initiation within days 0--29 or 0--89, and no initiation
before day 90, after which initiation was allowed) were analysed in 10-day
intervals. We took the tutorial's artificial-censoring model, with age, sex,
end-stage renal disease and, in the phase in which initiating causes
censoring, interval-number terms, as the pre-specified model.
The eleven monitored history functions were the five baseline covariates
(age, sex, renal disease and two race indicators), the three use
indicators and the three depth functions, which remained outside every
model and test whether the 10-day indicator summarises the dispensing
history. No function was designated as primary in advance, so the
function-specific comparisons are exploratory (Section~\ref{sec:rec}).
Cumulative RRB was computed for the five baseline covariates, with 95\%
percentile intervals of the individual bootstrap (200 resamples), and the
null reference had 500 draws. The associations of the use indicators with
initiation and with the outcome were estimated by pooled logistic
regression in the individual-intervals at risk under no initiation before
day 90, given age, sex, renal disease and interval. Following
the usual convention, and unlike Section~\ref{sec:notation}, censoring for
non-initiation was realised at the end of the last interval of a window,
after its outcomes and losses to follow-up, whose individuals stayed in
the risk set with $\hat\pi=1$; this keeps the zero targets of interval and
cumulative RRB and removes that of censoring balance (Appendix S1).
The case study therefore does not report censoring balance. Loss to
follow-up was rare (30 of the 1,590 patients before day 180, 8 by day 90)
and was not weighted.
By the identity of Section~\ref{sec:strategy}, a lost share of 0.005 and
survivor-versus-baseline differences of at most 0.17 bound its
contribution to cumulative RRB at about 0.001, if the lost patients
resemble the survivors and the weights do not concentrate on them.

\begin{table*}[tp]
\centering
\caption{Webster-Clark cohort: diagnostics and 180-day risks with the
tutorial's censoring model (age, sex, end-stage renal disease; base) and
with the three healthcare-use indicators added (revised).}
\label{tab:wcrace}
\footnotesize
\setlength{\tabcolsep}{4pt}
\resizebox{\textwidth}{!}{\begin{tabular}{@{}lll@{}}
\toprule
Quantity & Base & Revised \\
\midrule
Interval RRB, dispensing, day 30 (days 0--29) & 0.245 & 0.018 \\
Interval RRB, dispensing, day 90 (days 0--89) & 0.190 & 0.000 \\
Interval RRB, preceding-window indicator, day 90 (days 0--89) & 0.159 & 0.132 \\
Interval RRB, dispensings in 30 days, day 90 (days 0--89) & 0.195 & 0.077 \\
Interval RRB, dispensings since discharge, day 90 (days 0--89) & 0.178 & 0.101 \\
Interval RRB, race other, day 90 (days 0--89) & $-$0.112 & $-$0.110 \\
Cumulative RRB, race other, day 90 (days 0--89) & $-$0.087 & $-$0.085 \\
\begin{tabular}[t]{@{}l@{}}Cumulative RRB, five baseline covariates,\\
largest $|\cdot|$ over days 0--90 (no initiation)\end{tabular} & 0.015 & 0.015 \\
Studentised max., dispensing, days 0--29 & 2.45 / 1.97 (0.020) & 0.51 / 2.12 (0.573) \\
Studentised max., dispensing, days 0--89 & 2.20 / 2.00 (0.024) & 0.01 / 1.94 (0.994) \\
Studentised max., dispensing, no initiation & 2.68 / 2.72 (0.056) & 1.68 / 2.63 (0.597) \\
Studentised max., dispensings in 30 days, days 0--89 & 2.16 / 1.90 (0.024) & 0.99 / 1.91 (0.335) \\
Studentised max., dispensings in 30 days, no initiation & 3.04 / 2.79 (0.026) & 2.42 / 2.71 (0.118) \\
Studentised max., eleven functions, days 0--29 & 2.45 / 3.08 (0.192) & 1.24 / 2.96 (0.838) \\
Studentised max., eleven functions, days 0--89 & 2.20 / 3.06 (0.251) & 1.89 / 3.14 (0.385) \\
Studentised max., eleven functions, no initiation & 3.04 / 3.65 (0.216) & 2.44 / 3.61 (0.707) \\
ESS/$|\mathcal R|$ at day 90 (days 0--89) & 0.105 & 0.095 \\
99th weight percentile at day 90 (days 0--89) & 6.6 & 8.7 \\
\addlinespace
Risk, initiate within days 0--29 & 0.414 [0.332, 0.496] & 0.418 [0.328, 0.508] \\
Risk, initiate within days 0--89 & 0.378 [0.323, 0.432] & 0.373 [0.321, 0.426] \\
Risk, no initiation before day 90 & 0.360 [0.332, 0.389] & 0.361 [0.333, 0.389] \\
Days 0--89 vs.\ days 0--29 & $-$0.036 [$-$0.132, 0.060] & $-$0.045 [$-$0.140, 0.051] \\
\bottomrule
\end{tabular}}

\medskip
\raggedright
RRB weighted; day 30 and 90: the intervals ending at these days (grace ends
of the initiation strategies); cumulative RRB at the end of the grace
period and, for the five baseline covariates, the largest absolute value
over the nine censoring intervals of no initiation. Studentised max.: largest $|z|$ of the strategy over its intervals
with artificial censoring (one for each initiation strategy; nine for no
initiation, eight for the preceding-window indicator and the count since
discharge); entries, observed / null 95th percentile ($p$), 500 draws.
Dispensing: a Part D dispensing of a drug other than the index antiplatelet
in the 10 days before the interval; use indicators: this, a physician claim
and a hospital outpatient claim. Depth functions (preceding-window
indicator, dispensings in 30 days, since discharge): in no model. Eleven
functions: five baseline covariates, three use indicators, three depth
functions. Risks: 95\% CIs from 200 bootstrap resamples (normal
approximation); a direct port of the tutorial's continuous-time code gave
0.404 and 0.357 (days 0--29 and 0--89). CI, confidence interval; ESS,
effective sample size; RRB, risk-set reconstruction balance.
\end{table*}

\begin{figure*}[tp]
\centering
\includegraphics[width=0.9\textwidth]{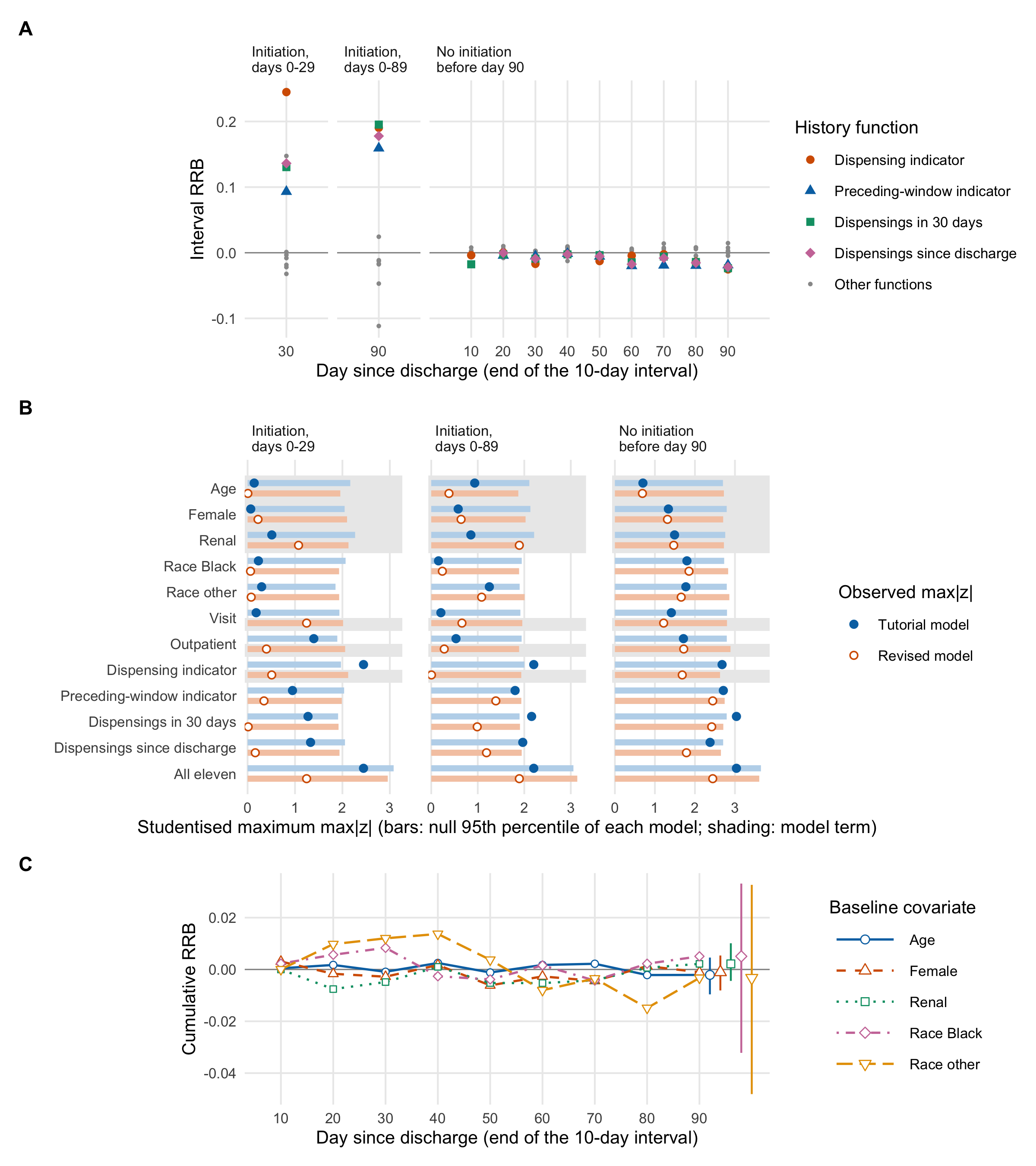}
\caption{RRB in the Webster-Clark cohort ($n=1{,}590$) under
the tutorial's censoring model (age, sex, end-stage renal disease) and the
revised model (plus the three use indicators). (A) Interval RRB of the
eleven history functions, tutorial model, one point per strategy and
censoring interval: coloured symbols, the dispensing indicator and its
three depth functions; grey, the other functions. (B) Studentised
maximum of interval RRB per strategy and history function (last row, all
eleven functions): filled blue circles, tutorial model (upper); open
vermillion circles, revised model (lower); bars, null 95th percentile (500
draws); shading, terms of the model on that half of the row. (C) Cumulative
RRB of the five baseline covariates under no initiation, tutorial model
(revised model within 0.003); whiskers,
bootstrap 95\% percentile intervals at day 90 (200 resamples). Renal,
end-stage renal disease; Visit, Outpatient and Dispensing indicator, a
physician claim, a hospital outpatient claim and a dispensing in the
preceding 10 days.}
\label{fig:wc}
\end{figure*}

\subsection{Results}
\label{sec:empres}
The 180-day risks were 0.414 and 0.378 for initiation within days 0--29 and
0--89 (risk difference $-0.036$ [$-0.132$, $0.060$]), close to those from
the tutorial's code (0.404 and 0.357; Table~\ref{tab:wcrace}). Under the
two initiation strategies, 83\% of the risk set was censored in days 20--29
(hereafter day 30) and 60\% in days 80--89 (hereafter day 90); under no
initiation before day 90, 4\% to 7\% was censored in each interval.
Weighting reduced $|$RRB$|$ of the modelled
covariates below 0.03 at day 30 and day 90 (Table S7). Cumulative RRB of
the five baseline covariates stayed within $\pm 0.015$ over the nine
censoring intervals, and at day 90 every 95\% bootstrap interval included
zero (Figure~\ref{fig:wc}C).

By day 90, 25.0\% of the cohort had the outcome. For end-stage renal
disease, the most prognostic baseline covariate, the depletion term in the
interval beginning on day 90 was $-0.086$ to $-0.071$ under the three
strategies, in agreement with the approximation of
Section~\ref{sec:strategy} ($-0.082$), whereas cumulative RRB was $-0.012$
to 0.002. The end-of-grace SMD was $-0.006$ between the two initiation
strategies, whose depletion terms cancelled, and $-0.024$ and $-0.030$
against no initiation.

A dispensing in the preceding 10 days predicted initiation in the interval
(odds ratio 1.37) and not the outcome (1.12), whereas the physician and outpatient indicators predicted
the outcome (1.44 and 1.40) and not initiation. Outside the censoring
model, the dispensing indicator kept an interval RRB of 0.245 at day 30 and
0.190 at day 90 after weighting (Figure~\ref{fig:wc}A). Its studentised
maximum exceeded the null 95th percentile under both
initiation strategies ($p=0.020$ and 0.024; Figure~\ref{fig:wc}B) and fell
just short of it under no initiation before day 90 ($p=0.056$). The three
depth functions had RRB of 0.159 to 0.195 at day 90, and the 30-day count
exceeded its reference under two strategies ($p=0.024$ and 0.026). The
maximum over all eleven functions stayed within its reference under every
strategy (Table~\ref{tab:wcrace}). Race other than white or Black
(prevalence 6\%, outside the model) had an interval RRB of $-0.112$ at day
90 under initiation within days 0--89 and stayed within its reference
($p=0.218$), but an ESS/$|\mathcal R|$ of 0.105 limits what this
non-exceedance shows.

Adding the three use indicators to both phases of the censoring model
reduced the interval RRB of the dispensing indicator to 0.018 at day 30
and 0.000 at day 90. The studentised maxima of the dispensing indicator,
of the depth functions (RRB 0.077 to 0.132 at day 90; $p\ge0.116$) and
over all eleven functions then stayed within their references
(Figure~\ref{fig:wc}B and Table~\ref{tab:wcrace}); race other stayed at
$-0.110$. At day 90 under initiation within days 0--89,
ESS/$|\mathcal R|$ fell from 0.105 to 0.095, the 99th weight percentile
rose from 6.6 to 8.7, and the risk difference moved from $-0.036$ to
$-0.045$ (Table~\ref{tab:wcrace}).

The evidence of misspecification rests on 4 of the 33 single-function
comparisons within a strategy ($p$ from 0.020 to 0.026; two more had
$p=0.050$), all concerning the dispensing history; with no function
designated in advance, these exploratory findings identify a candidate
revision. As the score equations nearly balance modelled functions
(Section~\ref{sec:interval}), the evidence that the revision improved the
reconstruction rests on the depth functions, which stayed outside both
models, and cumulative RRB was unchanged. The revised model was selected
with diagnostics from the same data, and its confidence interval does not
reflect this selection. We report it as the primary analysis of this
illustration and the tutorial's model as a sensitivity analysis; under
either, ESS/$|\mathcal R|$ of about 0.1 at day 90 leaves the estimate for
initiation within days 0--89 dependent on few retained individuals.
\section{Discussion}
\label{sec:disc}

\subsection{Principal findings}
Risk-set reconstruction balance measures how far the weighted retained
individuals depart from the risk set just before censoring (interval RRB)
and from the baseline population (cumulative RRB), the two populations
that correct weights reconstruct (Propositions~1 and~2). In the
simulation, interval RRB responded to an omitted term's effect on
censoring whether or not the term was prognostic. Under correct models,
the exceedance rate of its studentised maximum was near the nominal level
when the censoring model pooled nine intervals and below it under
censoring in a single interval. Unstable weights lowered the coverage of
$\hat\theta_{32}$ to between 0.500 and 0.650 without raising the exceedance rate,
so weight instability is assessed with the ESS and the weight percentiles.
With many early events, cumulative RRB stayed near zero under correct
weights, whereas the end-of-grace SMD departed from zero through the
depletion term; under loss to follow-up that the weights ignored,
cumulative RRB departed from zero and interval RRB stayed small. In the
case study, exploratory comparisons flagged the dispensing indicator, and
after the revision the functions outside the model stayed within their
references, cumulative RRB was unchanged and the weights became more
concentrated.

\subsection{Relation to previous work}
The propositions follow from standard properties of inverse probability
weighting. The contribution lies in assigning each diagnostic in use its
reference population and the conditions under which it has a zero target.
Jackson defines censored-versus-retained balance measures for loss to follow-up and for
artificial censoring at protocol deviation \cite{jackson2019}. By
Proposition~1 the measure for artificial censoring has the same zero
target as interval RRB. The balance tables of cobalt and WeightIt
\cite{greifer2026cobalt,greifer2026weightit} compare the retained
individuals, weighted by the product of the weights of all listed models,
with the unweighted risk set, standardised by its unweighted standard
deviation. This comparison coincides with interval RRB if two conditions
hold. The earlier weights $W_{k-1}$ are constant in the risk set, as in
the first interval with censoring, and the weights of the retained
individuals include no censoring model of a later interval. Otherwise, the
reference lacks the earlier weights or the weights of the retained
individuals include later intervals, and the comparison has in general no
zero target under a correct censoring model.
Calibrated censoring weights \cite{yiu2022} impose balancing conditions
of the same form during estimation, and covariate balancing propensity
scores \cite{imai2014} impose analogous conditions on treatment weights.
If conditions of this form hold exactly in the same
stratum and against the same reference as RRB, the RRB of the functions
they contain is zero by construction, and the diagnostic information lies
in the functions outside the conditions.
The CCW weights standardise the retained individuals to the baseline
covariate distribution \cite{gaber2024b}, as per-protocol weights make
adherence independent of the measured prognostic factors
\cite{hernan2017}; a comparison with the baseline population restricted
to survivors carries the depletion term of (\ref{eq:deplete}).

The null reference proposed here judges RRB against its chance variation
under the fitted censoring model and therefore resembles a balance test
\cite{hansen2008,hripcsak2025}.
Imai, King and Stuart argue that balance tests are inappropriate stopping
rules for matching, because balance is a property of the sample and test
results depend on the sample size \cite{imai2008}. The null reference
tests the specification of the censoring model, not the balance of the
sample, as the test of over-identifying restrictions for covariate
balancing propensity scores tests the specification of the propensity score
model \cite{imai2014}. The power of the null reference still depends on
the sample size, so RRB is reported as a descriptive quantity, together
with the null reference and the ESS.

\subsection{Limitations and future work}
Imbalance and bias need not agree. The diagnostics can flag imbalance in a
function that is not prognostic, and a small RRB does not establish that
the effect estimate is unbiased, for example when an omitted covariate is
strongly correlated with modelled functions. Detection depends on which
history functions are monitored, and no set of functions guarantees it. In
strata with few censored individuals or a small ESS, a few weights
dominate the diagnostics. The calibration of the null reference is
approximate. Its exceedance rate was below the nominal level for
strategies with a single censoring interval and could exceed it when few
censoring events support many model parameters; a calibrated variant
remains future work. Cumulative RRB has no null reference, as one would
require models for the covariate and outcome histories after censoring.
The simulation had at most nine baseline covariates, one time-varying
covariate, absorbing initiation and 12 intervals, and the case study
involved an omission known in advance. More time-varying covariates, more
severe violations of positivity and sustained strategies remain to be
studied.

\subsection{Recommendations}
\label{sec:rec}
We recommend reporting three checks, each for a different failure
mechanism. (i) Misspecification of the artificial-censoring model is
checked by interval RRB or censoring balance, with the studentised maximum
of each strategy against the null reference. (ii) Loss of representation
of the baseline population, through loss to follow-up that the weights
ignore or weight truncation, is checked by cumulative RRB of baseline
covariates up to the end of the grace period, reported as a descriptive
trajectory. (iii) Weight concentration is checked by the ESS and weight
percentiles by stratum. The end-of-grace SMD has no zero target once
events have occurred, and between-strategy SMDs of post-baseline
covariates have none.

As the score equations nearly balance the functions in the model, the
monitored set should include history functions outside it. A few of these
should be specified in advance and compared with their own null
references, and further comparisons are exploratory. An exceedance prompts
a revision of the specification, which is judged by the three checks
together, and the effect estimate is reported under both models.

\subsection{Conclusion}
In CCW analyses under positivity and a correct censoring model, the
weighted individuals retained under a strategy reconstruct the risk set before each censoring
interval and the baseline population. Risk-set reconstruction balance
measures how far the weighted retained individuals depart from these two
populations. Interval RRB has a zero
target for every history function and, judged against the null reference
for chance imbalance, can serve as a check of the censoring model.
Cumulative RRB keeps a zero target for baseline covariates after outcome
events and can serve as a check of the representation of the baseline
population.

\section*{Funding}
This work was supported by JST ASPIRE, Japan [grant number JPMJAP2403]. The
funder had no role in the study design; in the collection, analysis and
interpretation of data; in the writing of the report; or in the decision to
submit the article for publication.

\section*{Conflict of interest}
The author declares no potential conflict of interests.

\section*{Data availability statement}
The R package ccwbalance (version 0.1.0), the code and the definitions of
the simulation scenarios that reproduce every table and figure, and the
saved simulation outputs they read are available in the GitHub repository
\url{https://github.com/Nonokyu/ccwbalance} and archived at Zenodo
(doi: 10.5281/zenodo.24013295).
The Webster-Clark SynPUF cohort is public and available with the tutorial code
at \url{https://github.com/mawcpharmdphd/CCW_tutorial}, and the DE-SynPUF
claims are public at the website of the Centers for Medicare \& Medicaid
Services. The repository of this work contains the scripts that extract
the cohort's claims, derive the time-varying indicators and re-create the
empirical outputs.

\section*{Software}
The diagnostics are implemented in the R package ccwbalance, version 0.1.0
(this work), developed and run in R 4.4.3, and the repository reproduces
every table and figure. The package also handles sustained strategies, in
which artificial censoring can occur in every interval. Appendix S4
describes its handling of edge cases and weight truncation.

\section*{Supporting Information}
Appendix S1: propositions and proofs. Appendix S2: simulation study,
with the data-generating mechanisms of Parts 1--3, the coupling of the null
reference within individuals, a per-strategy decomposition of its
exceedance shortfall (Table S1), the dependence of MaxIB on the scale of
the history functions, robust SE ratios under correct specification,
null-reference exceedance by history function (Table S2), performance and
diagnostics of Parts 1--3 (Table S3), detection by interval RRB and
censoring balance at nominal and at matched empirical size (Table S4),
factorial contrasts of Part 1 (Table S5) and bootstrap
intervals of cumulative RRB (Table S6). Appendix S3:
case study, with interval RRB at the grace-period ends (Table S7). Appendix S4: implementation details of the package.

\bibliographystyle{vancouver-doi}
\bibliography{references}

\end{document}